\documentclass{mem}
\usepackage{natbib}\usepackage{txfonts}\usepackage{balance}
\usepackage{graphicx}
\usepackage[a4paper,breaklinks,dvipdfm]{hyperref}
\idline{75}{282}
\begin{document}
\def\teff{$T\rm_{eff }$}
\def\kms{$\mathrm {km s}^{-1}$}

\title{
Theoretical Modeling of the RR Lyrae Variables in NGC$\,$1851
}

   \subtitle{}

\author{
A. \,Kunder\inst{1} 
M. Salaris\inst{2},
S. Cassisi\inst{3}, 
R. de Propris\inst{1},
A. Walker\inst{1},
P. B. Stetson\inst{4},
M. Catelan\inst{5,6},
\and
P. Amigo\inst{5,6}
 }

\offprints{A. Kunder}

\institute{
NOAO-Cerro Tololo Inter-American Observatory, Casilla 603, La Serena, Chile
\and
Astrophysics Research Institute, Liverpool John Moores University, Twelve Quays House, Egerton Wharf, Birkenhead CH41 1LD, UK
\and
INAF-Osservatorio Astronomico di Collurania, Via M. Maggini, I-64100 Teramo, Italy
\and
Dominion Astrophysical Observatory, Herzberg Institute of Astrophysics, National Research Council, Victoria BC, Canada
\and
Pontificia Universidad Cat\'olica de Chile, Departamento de Astronom\'\i a y Astrof\'\i sica, Av. Vicu\~{n}a Mackenna 4860, 782-0436 Macul, Santiago, Chile
\and
The Milky Way Millennium Nucleus, Av. Vicu\~{n}a Mackenna 4860, 782-0436 Macul, Santiago, Chile
}

\authorrunning{Kunder }

\titlerunning{RR Lyrae Variables of NGC$\,$1851}

\abstract{
The RR Lyrae instability strip (IS) in NGC$\,$1851 is investigated, and a model is presented 
which reproduces the pulsational properties of the RR Lyrae population.  In our model,
a stellar component within the IS that displays minor 
helium variations ($Y$$\sim$0.248-0.270) is able to reproduce the observed periods and 
amplitudes of the RR Lyrae variables, as well as the frequency of fundamental and 
first-overtone RR Lyrae variables.  The RR Lyrae variables therefore may belong to an 
He-enriched second generation of stars.  The RR Lyrae 
variables with a slightly enhanced helium ($Y$$\sim$0.270-0.280) have longer periods at a 
given amplitude, as is seen with Oosterhoff II (OoII) RR Lyrae variables, whereas the 
RR Lyrae variables with $Y$$\sim$0.248-0.270 have shorter periods, exhibiting properties 
of Oosterhoff I (OoI) variables. %
%\keywords{Stars: Population II -- Galaxy: globular clusters }
}
\maketitle{}

\section{Introduction}
The globular cluster (GC) NGC$\,$1851 hosts two distinct subgiant 
branches (SGBs) \citep{milone08}, prompting much discussion as to the formation 
history of this cluster.  The split between the bright SGB (SGBb) and faint SGB (SGBf) can 
be explained by two subpopulations that differ in age by 
about 1 Gyr \citep[e.g.,][]{milone08, gratton12}, or by
%about 1 Gyr \citep[e.g.,][]{milone08, carretta11a, carretta11b, gratton12}, or by
two SGBs that are nearly coeval but have different C+N+O contents \citep[e.g.,][]{cassisi08, ventura09}.  
The horizontal branch (HB) of NGC$\,$1851 also hosts two distinct populations, namely a 
prominent red HB clump and a blue tail.  Finally, different populations have also been 
revealed in the red giant branch (RGB) \citep[e.g.,][]{grundahl99, han09}.  
%revealed in the red giant branch (RGB) \citep[e.g.,][]{grundahl99, calamida07, lee09, han09}.  

The HB of globular clusters (GCs) is a valuable stellar component which
can be used to investigate the formation and evolution of GCs \citep[e.g.][]{gratton10, dotter10}.
Indeed, there have been previous studies to model the HB of the NGC$\,$1851 to 
obtain scenarios of the formation this cluster's bimodal horizontal branch \citep{salaris08, gratton12}.
Our study concentrates on the portion of the
HB inside the instability strip (IS), by discussing the case of the RR Lyrae properties
in NGC$\,$1851.  
%We incorporate the many detailed spectroscopic and photometric investigations of the 
%stellar population of NGC$\,$1851, which eliminates the need to adopt many assumptions that contribute
%to the particular HB morphologoy of a GC.   

\section{RR Lyrae Observations }
Because of the compact nature NGC$\,$1851, it is likely that the RR Lyrae star sample
in the crowded core in incomplete \citep[e.g.][]{sumerel04, downes04}.  
%For example, recently \citet{sumerel04} discovered 19 variables 
%and \citet{downes04} reported eleven new variables within 
%40" of the cluster center.  Unfortunately, due to crowding issues,
%placing these inner RR Lyrae light curves on a photometrically calibrated scale was not
%possible.  
Fortunately at distances greater than 40" from the cluster center, individual stars 
can be relatively easily resolved, and there is no indication that the RR Lyrae sample is 
incomplete in this region.  Therefore we limit our sample of RR Lyrae stars with which 
to compare our HB models to this outer region.  

\begin{figure}[]
\resizebox{\hsize}{!}{\includegraphics[clip=true]{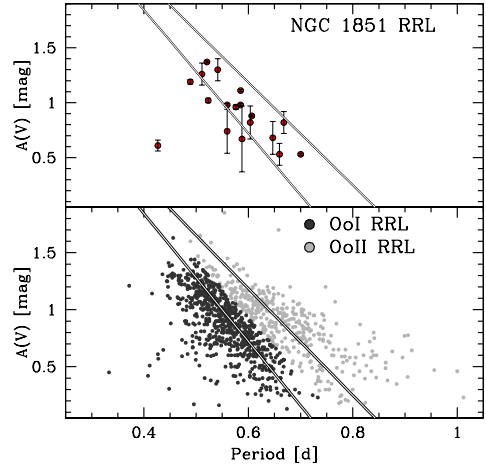}}
\caption{
\footnotesize
 {\it Top:} Period-amplitude diagram for our sample of RR0 Lyrae variables in NGC$\,$1851.
{\it Bottom:} Period-amplitude relation for 1097 RR0 Lyrae variables in 40 Galactic GCs.
Our division of OoI and OoII-type RR Lyrae variables are shown
by dark and light circles, respectively.
The lines derived by \citet{clement99a} for OoI and OoII RR0 are overplotted.  }
\label{PAall}
\end{figure}

The outer RR Lyrae variables were studied by \citet{walker98} using the $BVI$
passbands, and robust mean magnitudes, periods and amplitudes are available.
The periods and $V$-amplitudes of our sample of fundamental mode RR Lyrae 
 (RR0) variables are shown in
Figure~\ref{PAall}, and the period-amplitude relation of typical
OoI and OoII-type systems is over-plotted.
We note that although many of the RR Lyrae stars in NGC$\,$1851 have periods and 
amplitudes that cause them to fall near the OoI PA relation, there 
are a number of stars following the OoII PA relation.

%Scatter in the PA plane can be introduced by the Blazhko effect, or other
%effects such as a rapidly changing period \citep{clement99a}.  Because
%the \citet{walker98} RR Lyrae variables were observed over a long time frame (126 total frames
%observed over 15 nights during a 1.5 year time span), a visual determination of the 
%change in amplitudes in each RR0 can be obtained.  These are shown as an error-bar in Figure~\ref{PAall}.  

In comparison, Figure~\ref{PAall} shows the periods and $V$-amplitudes of 1097 RR0 
Lyrae variables in 40 Galactic globular clusters \citep{kunder13}.
This sample of RR0 Lyrae variables is divided by their position in the period-amplitude plane following
the lines that \citet{clement99a} derived for Oosterhoff I and Oosterhoff II RR0 stars (see
Figure~\ref{PAall}).  Those RR Lyrae variable falling closest to the OoI line are designated as 
OoI-type RR Lyrae stars, and those falling closest to the OoII line are the OoII-type variables.
  
\begin{figure}[htb]
\includegraphics[width=1\hsize]{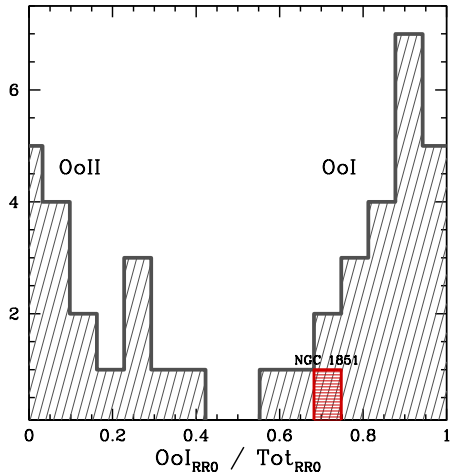}
\caption{A histogram of the ratio of OoI stars in the Milky Way GCs.  The Oosterhoff ratio of the
NGC$\,$1851 RR0 Lyrae stars is highlighted.  
\label{ooratio}}
\end{figure}

We define an Oosterhoff ratio for each GC, which is simply the number of OoI-type RR0 Lyrae stars
compared to the total number of RR0 Lyrae stars in the GC, $\rm OoI_{RR0} / Tot_{RR0}$.
Figure~\ref{ooratio} shows the histogram of the Oosterhoff ratio of the GCs in our sample.
%Many of these 40 GCs lack a complete sample of RR Lyrae variables and it is certain that
%at least some of the RR Lyrae amplitudes are affected by the Blazhkocity or other
%light curve ``noise".  However, despite these caveats, 
Our defined Oosterhoff ratio splits the Milky Way GCs into two groups; the 
OoI-type clusters have RR Lyrae variables with shorter periods for a given 
amplitude and hence have larger Oosterhoff ratios (with respect to the OoII-type
clusters).  Further, there is an absence of clusters 
falling in the ``gap".  We therefore believe that our Oosterhoff ratio is useful
to distinguish between OoI- and OoII-type GCs.  

\section{Synthetic HB Modeling} 
Our synthetic HB calculations are carried out to provide
a simple and attractive explanation for the cluster HB and IS morphology, 
keeping the number of free parameters to a minimum, yet still reproducing the RR Lyrae
properties that make this cluster stand out as having an unusual Oosterhoff type.
The HB evolutionary tracks used are from the BaSTI stellar 
library \citep{pietrinferni04, pietrinferni06, pietrinferni09}.
%and have been 
%employed previously to model the HB of this cluster \citep{salaris08, cassisi08, gratton12}.   
The horizontal part of the HB, which includes 
the RR Lyrae instability strip, makes up $\sim$10\% of the cluster
stellar sample brighter than $V$ = 21 mag; this is the component our study focuses on modelling.  
To be considered a match to the observations, the synthetic HB model must 
agree with four specific observational constraints.
The first is that the number ratio of the progeny of the SGBb to the progeny of the SGBf 
subpopulations has to satisfy the 70:30 ratio observed along the SGB \citep{milone09}. 
%This population ratio takes into account the possible change in the evolutionary rate between 
%SGBb and SGBf stars due to the differences in their chemical abundances \citep[see][for details]{milone09}.
The second is that the
$\rm (B\thinspace{:}V\thinspace{:}R)$ (blue\thinspace{:}variable\thinspace{:}red HB) 
ratio agrees with the observed 
$\rm (B\thinspace{:}V\thinspace{:}R)$ = $\rm (33\pm8\thinspace{:}10\pm5\thinspace{:}56\pm11)$
\citep[in line with the results by][]{catelan98, saviane98}.
The third requirement is that the theoretical periods and amplitudes agree with those
observed, and the fourth that the frequency of first-overtone RR Lyrae stars (RR1) to 
RR0 variables is approximately 
$N_{\rm 1}/N_{\rm tot}$ $\sim$ 0.35 \citep{walker98, downes04}.  

The four HB components described by \citet{gratton12} are used as a starting
points for our calculations.  As in \citet{gratton12}, The only parameter we vary is the initial He abundance of the HB 
progenitors, keeping the same total RGB mass loss for all components.  That the mass loss
is the same for all components is not something we can prove, and we stress that this is our 
working assumption, and our results follow from that. 

\begin{figure}[htb]
\includegraphics[width=1\hsize]{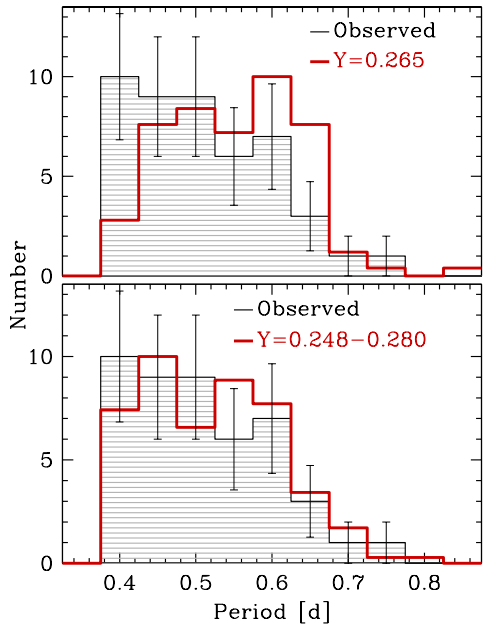}
\caption{
{\it Top:}  
A comparison between the observed NGC$\,$1851 RR Lyrae periods versus the theoretical 
periods with a constant Y=0.265 (Gratton et~al. 2012).  
{\it Bottom:}  The results when using
a He abundance with minor helium variations ($Y$$\sim$0.248-0.280) versus the observed periods.  
}
\label{PAhist}
\end{figure}

Objects from our synthetic HB that fall within the observed IS from \citet{walker98} are
considered RR Lyrae variables.
\citet{gratton12} model the horizontal part of 
the HB component with a constant He abundance $Y$=0.265. 
We find a much better fit to the RR Lyrae properties, however, by assuming
a continuous He distribution between $Y$=0.248 and 0.280 (see Figure~\ref{PAhist}). 
In our simulation, the mean He abundance in the IS is $<Y>$=0.271, close to 
the constant abundance $Y$=0.265 employed by \citet{gratton12} for this component,
and the mean mass is ${\rm <M>}$ = $\rm 0.634~M_{\sun}$. 

Figure~\ref{PAhist} shows the periods of the RR Lyrae variables in our simulation
as compared to the observed periods, where the theoretical periods are calculated 
for all HB stars falling within the observed IS using the
\citet{dicriscienzo04} RR Lyrae pulsation models.  The RR1 variables are fundamentalized via 
log $\rm P_0 \sim log P + 0.127$, where $\rm P_0$ is the fundamental mode period.
A Kolmogorov-Smirnov (KS) test between the observed periods and the synthetic RR Lyrae periods
returns a probability P=0.86, well above the default threshold 
${\rm P_{th}}$=0.05 below which one rejects the null hypothesis.
On this basis, we find that the synthetic
periods from our simulated HB and the observed periods agree well 
with each other. 

The ratio of first overtone to total RR Lyrae variables in our simulation is
$N_{\rm 1}/N_{\rm tot}$ $\sim$0.30, in agreement with that observed.
In contrast, simulations using a constant helium for the portion of the HB containing 
the IS \citep[as in][]{gratton12} do not fit the constraints given by the NGC 1851 RR Lyrae
variables as well. For example, adopting $Y$=0.265 results in an $\rm N_1/N_{tot}$ = 0.11 
(versus the observed $\rm N_1/N_{tot}$ = 0.28).

\section{Conclusions}
Modern theoretical synthetic HB models are used to investigate the RR Lyrae properties
of NGC$\,$1851.  
%Thanks to the large number of recent papers 
%devoted to this cluster with high quality observations, very few input parameters in 
%our models are unknown.  
Both the RR Lyrae period distribution and the number 
ratio of first overtone RR Lyrae to total RR Lyrae stars, $\rm N_1/N_{tot}$, provide constraints 
pertaining to the component of the HB containing the IS.  It is straightforward to reproduce 
the observed distribution of RR Lyrae stars inside the instability strip with minor 
He variations ($Y$$\sim$0.248-0.280). 

In general, the RR Lyrae variables with $Y$ $<$ 0.27 fall in the OoI area of the PA diagram, 
whereas the RR Lyrae variables with $Y$ $>$ 0.27 fall close to the OoII line.  
Assuming that the period-amplitude diagram can be effectively used to classify RR Lyrae 
stars into an Oosterhoff type, this means that He and Oosterhoff type are correlated
in this cluster.  This is not completely unexpected, as an increase in He makes RR Lyrae 
variables brighter and, as a consequence, higher helium abundance makes their 
pulsational period longer at fixed temperature/color \citep{bono97}.

%The pulsation periods and 
%amplitudes from the RR Lyrae variables resulting from variations in He along the IS
%have different characteristics.  The RR Lyrae variables with a ``normal" helium
%have periods and amplitudes, as well as a $N_{\rm 1}/N_{\rm tot}$
%ratio, that is inline with OoI-type GCs.  In contrast, the RR Lyrae variables with slightly enhanced
%He (0.27 $<$ $Y$ $<$ 0.28) have longer periods and a higher ratio of $N_{\rm 1}/N_{\rm tot}$, 
%indicative of RR Lyrae variables in OoII-type GCs.  
%In the absence of spectroscopy of the RR Lyrae variables
%in NGC$\,$1851, the synthetic horizontal part of the HB and RR Lyrae instability strip 
%presented here is the simplest one that reproduces the 
%available observations with the smallest amount of free parameters.  New observations
%of the RR Lyrae variables may require more complex modeling, however, and would be
%particularly interesting.

\bibliographystyle{aa}

\end{document}